\documentclass{ws-mpla}

\begin{document}

\markboth{S.~Scherer} {Effective Field Theory of the Single-Nucleon
Sector}

\catchline{}{}{}{}{}

\title{EFFECTIVE FIELD THEORY OF THE SINGLE-NUCLEON SECTOR}

\author{\footnotesize STEFAN SCHERER}

\address{Institut f\"ur Kernphysik, Johannes
Gutenberg-Universit\"at, J.~J.~Becher-Weg 45\\
D-55099 Mainz, Germany\\
scherer@kph.uni-mainz.de}
\maketitle

\pub{Received (Day Month Year)}{Revised (Day Month Year)}

\begin{abstract}
We address the issue of a consistent power counting scheme in
manifestly Lorentz-invariant baryon chiral perturbation theory. We
discuss the inclusion of vector mesons in the calculation of the
nucleon electromagnetic form factors. We comment on the chiral
expansion of the nucleon mass to order ${\cal O}(q^6)$.

\keywords{Chiral lagrangians; renormalization.}
\end{abstract}

\ccode{PACS Nos.: 12.39.Fe, 11.10.Gh}

\section{Introduction}
   Effective field theory (EFT) is a powerful tool for describing the
strong interactions at low energies \cite{Weinberg:1978kz}.
  Starting point is the chiral $\mbox{SU}(N)_L\times\mbox{SU}(N)_R$
symmetry of QCD in the limit of $N$ massless quarks and its
spontaneous breakdown to $\mbox{SU}(N)_V$ in the ground state.
   Instead of solving QCD in terms of quarks and gluons,
its low-energy physics (of the mesonic sector) is described using
the most general Lagrangian containing the Goldstone bosons as
effective degrees of freedom
\cite{Gasser:1983yg,Gasser:1984gg,Fearing:1994ga,%
Bijnens:1999sh,Ebertshauser:2001nj,Bijnens:2001bb}.
    Physical quantities are calculated in terms of
an expansion in $p/\Lambda$, where $p$ stands for momenta or masses
that are smaller than a certain momentum scale $\Lambda$ (see, e.g.,
Refs.~\cite{Scherer:2002tk,Scherer:2005ri} for an introduction).
  In the following we will outline some recent developments in devising
a renormalization scheme leading to a simple and consistent power
counting for the renormalized diagrams of a manifestly
Lorentz-invariant approach to baryon chiral perturbation theory
\cite{Gasser:1987rb}.

\section{Renormalization and Power Counting}
  The standard effective Lagrangian relevant to the single-nucleon sector
  consists of the sum of the purely mesonic and $\pi N$ Lagrangians,
respectively,
\begin{displaymath}
{\cal L}_{\rm eff}={\cal L}_{\pi}+{\cal L}_{\pi N}={\cal L}_2+ {\cal
L}_4 +\cdots +{\cal L}_{\pi N}^{(1)}+{\cal L}_{\pi N}^{(2)}+\cdots
\end{displaymath}
which are organized in a derivative and quark-mass expansion.
   The aim is to devise a renormalization procedure generating, after
renormalization, the following power counting:
   a loop integration in $n$ dimensions counts as $q^n$,
pion and fermion propagators count as $q^{-2}$ and $q^{-1}$,
respectively, vertices derived from ${\cal L}_{2k}$ and ${\cal
L}_{\pi N}^{(k)}$ count as $q^{2k}$ and $q^k$, respectively.
   Here, $q$ generically denotes a small expansion parameter such as,
e.g., the pion mass.

    Several methods have been suggested to obtain a consistent
power counting in a manifestly Lorentz-invariant approach.
   As an illustration consider the integral
\begin{displaymath}
H(p^2,m^2;n)= \int \frac{d^n k}{(2\pi)^n}
\frac{i}{[(k-p)^2-m^2+i0^+][k^2+i0^+]},
\end{displaymath}
where $\Delta=(p^2-m^2)/m^2={\cal O}(q)$ is a small quantity. In the
infrared (IR) regularization of Becher and Leutwyler
\cite{Becher:1999he} one makes use of the Feynman parametrization
\begin{displaymath}
{1\over ab}=\int_0^1 {dz\over [az+b(1-z)]^2}
\end{displaymath}
with $a=(k-p)^2-m^2+i0^+$ and $b=k^2+i0^+$.
   The resulting integral over the Feynman parameter $z$ is then rewritten as
\begin{eqnarray*}
\int_0^1 dz \cdots &=& \int_0^\infty dz \cdots
- \int_1^\infty dz \cdots,\\
\end{eqnarray*}
where the first, so-called infrared (singular) integral satisfies
the power counting, while the remainder violates power counting but
turns out to be regular and can thus be absorbed in counterterms.
 The central idea of the extended on-mass-shell (EOMS)
scheme\cite{Gegelia:1999gf,Fuchs:2003qc} consists of performing
additional subtractions beyond the $\widetilde{\rm MS}$ scheme.
  In Ref.\ \cite{Schindler:2003xv} the IR regularization of
Becher and Leutwyler was reformulated in a form analogous to the
EOMS renormalization scheme.
   Within this (new) formulation the subtraction terms are found by
expanding the integrands of loop integrals in powers of small
parameters (small masses and Lorentz-invariant combinations of
external momenta and large masses) and subsequently exchanging the
order of integration and summation.
   The new formulation of IR regularization can be applied to
diagrams with an arbitrary number of propagators with various masses
(e.g., resonances) and/or diagrams with several fermion lines as
well as to multi-loop diagrams.

\section{Applications}
\subsection{Nucleon Form Factors}
\begin{figure}[ph]
\centerline{\psfig{file=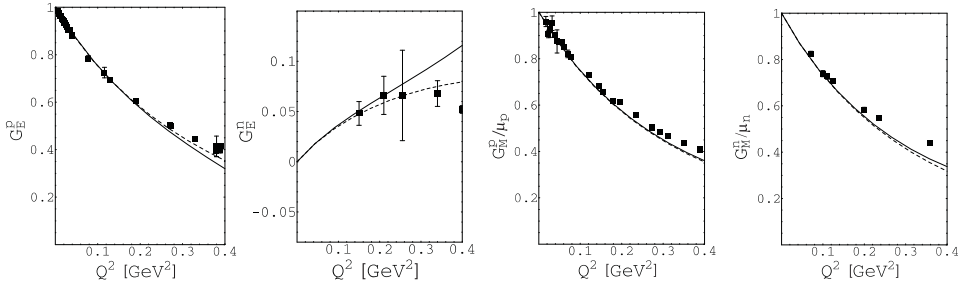,width=\textwidth}}
\vspace*{8pt} \caption{The Sachs form factors of the nucleon in
manifestly Lorentz-invariant chiral perturbation theory at ${\cal
O}(q^4)$ including vector mesons as explicit degrees of freedom.
Full lines: results in the extended on-mass-shell scheme; dashed
lines: results in infrared regularization.\protect\label{G_neu}}
\end{figure}

   It has been known for some time that ChPT  results at ${\cal O}(q^4)$
only provide a decent description of the electromagnetic Sachs form
factors $G_E$ and $G_M$ up to $Q^2=0.1\,\mbox{GeV}^2$ and do not
generate sufficient curvature for larger values of $Q^2$
\cite{Kubis:2000zd,Fuchs:2003ir}. To improve these results
higher-order contributions have to be included. This can be achieved
by performing a full calculation at ${\cal O}(q^5)$ which would also
include the analysis of two-loop diagrams.
   Another possibility is to include additional degrees of freedom, through which
some of the higher-order contributions are re-summed.
   Both the reformulated IR regularization and the EOMS scheme allow for
a consistent inclusion of vector mesons which already a long time
ago were established to play an important role in the description of
the nucleon form factors.
    Figure \ref{G_neu} shows the results for
the electric and magnetic Sachs form factors in the EOMS scheme
(solid lines) and the infrared renormalization (dashed lines)
\cite{Schindler:2005ke}.
   A {\em consistent} inclusion of vector
mesons clearly  improves the quality of the description.
   Similarly, the inclusion of the axial-vector meson $a_1(1260)$ results
in an improved description of the experimental data for the axial
form factor \cite{Schindler:2006it}.

\subsection{Chiral Expansion of the Nucleon Mass to Order ${\cal O}(q^6)$}

Using the reformulated infrared regularization
\cite{Schindler:2003xv} we have calculated the nucleon mass up to
and including order ${\cal O}(q^6)$ in the chiral expansion
\cite{Schindler:2006ha,Schindler:2007dr}:
\begin{eqnarray}\label{H1:emff:MassExp}
    m_N &=& m +k_1 M^2 +k_2 \,M^3 +k_3 M^4 \ln\frac{M}{\mu}
+ k_4 M^4  + k_5 M^5\ln\frac{M}{\mu} + k_6 M^5  \nonumber\\&& + k_7
M^6 \ln^2\frac{M}{\mu}+ k_8 M^6 \ln\frac{M}{\mu} + k_9 M^6.
\end{eqnarray}
   In Eq.~(\ref{H1:emff:MassExp}), $m$ denotes the nucleon mass in the
chiral limit, $M^2$ is the leading term in the chiral expansion of
the square of the pion mass, $\mu$ is the renormalization scale; all
the coefficients $k_i$ have been determined in terms of infrared
renormalized parameters.
   Our results for the renormalization-scheme-independent terms agree
with the heavy-baryon ChPT results of Ref.~\cite{McGovern:1998tm}.

\begin{figure}[ph]
\centerline{\psfig{file=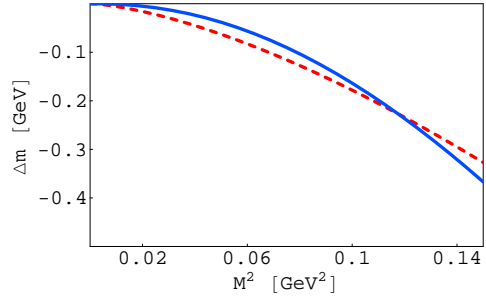,width=0.5\textwidth}}
\vspace*{8pt} \caption{Pion mass dependence of the term $k_5 M^5
\ln(M/m_N)$ (solid line) for $M<400$ MeV. For comparison also the
term $k_2 M^3$ (dashed line) is shown.
\protect\label{fig:nucleonmass}}
\end{figure}

  The numerical contributions from higher-order terms cannot be
calculated so far since, starting with $k_4$, most expressions in
Eq.~(\ref{H1:emff:MassExp}) contain unknown low-energy coupling
constants (LECs) from the Lagrangians of order ${\cal O}(q^4)$ and
higher.
   The coefficient $k_5$ is free of higher-order LECs.
   Figure \ref{fig:nucleonmass} shows the pion mass dependence of the term
$k_5 M^5 \ln(M/m_N)$ (solid line) in comparison with the term $k_2
M^3$ (dashed line) for $M<400$ MeV.
   For $M\approx 360\,\mbox{MeV}$ the $k_5$ term is as large as the
   $k_2$ term.

\section*{Acknowledgments}
  This work was
made possible by the financial support from the Deutsche
Forschungsgemeinschaft (SFB 443 and SCHE 459/2-1) and the EU
Integrated Infrastructure Initiative Hadron Physics Project
(contract number RII3-CT-2004-506078).

\end{document}